\documentclass{jfm}
\usepackage{float}
\usepackage{graphicx}
\usepackage{epstopdf, epsfig}
\usepackage{amsmath}
\usepackage{soul}
\usepackage{color}
\usepackage{siunitx}
\usepackage{subcaption}
\usepackage{commath}
\usepackage[dvipsnames]{xcolor}
\usepackage{xfrac}
\usepackage{cancel}
\usepackage{color, soul}

\newcommand\PFL[1]{{\color{black}#1}}
\newcommand\pfl[1]{{\color{black}#1}}

\newcommand\rkb[1]{{\color{black}#1}}

\newcommand{\TT}{\textsf{T}}

\newcommand{\ub}{\textbf{u}}

\newcommand{\nb}{\textbf{n}}
\newcommand{\Tb}{\textbf{T}}

\newcommand{\zb}{\textbf{z}}
\newcommand{\rb}{\textbf{r}}

\shorttitle{The circular capillary jump}
\shortauthor{R.K. Bhagat, P.F. Linden}

\title{The circular capillary jump}

\author{Rajesh K. Bhagat
  \corresp{\email{rkb29@cam.ac.uk}},
  and P. F. Linden,
 }

\affiliation{
Department of Applied Mathematics and Theoretical Physics,\\ Wilberforce Road, Cambridge CB3 0WA, UK}

\begin{document}

\maketitle

\begin{abstract}
In this paper we {re-}examine the flow produced by the normal impact of a laminar liquid jet onto an infinite plane {when the flow is dominated by surface tension.} {Over the range of parameters we consider, which are typical of water from a tap over a kitchen sink, }it is observed experimentally that after impact the liquid spreads radially over the plane away from the point of impact in a thin film. It is also observed that, at a finite radius, there is an abrupt increase in thickness of the film which has been identified as a hydraulic jump. {Once the jump is formed} this radius { remains constant in time and, further, }is independent of the orientation of the surface showing that gravity is unimportant \citep{bhagat2018origin}. 
We show that the application of conservation of momentum in the film, subject {only} to viscosity and surface tension {and ignoring gravity completely}, predicts a singularity in the {curvature of the liquid film and consequently a jump in} the depth of the film at a finite radius. This location is almost identical to the radius of the jump predicted by conservation of energy and agrees with experimental observations. 
\end{abstract}

\section{Introduction}
 \setlength{\belowcaptionskip}{-12pt}
\label{sec_introduction}

 In a recent paper \cite{bhagat2018origin} conducted experiments that showed that in a thin liquid film, on scales typical {of those found} in a kitchen sink, the {circular} jump produced by the normal impact of a {round} laminar jet onto an infinite plane is independent of the orientation of the surface. These experiments conclusively showed that gravity {does} {\it not} play a significant role in the origin/formation of these jumps -- in sharp contrast with previous theoretical analyses. 
 They also used conservation of energy, including both surface tension {\it and} gravity, 
 to determine the radius of the jump. As we discuss below, on the scale of a kitchen sink the predicted radius is found to be almost independent of gravity, and is in excellent agreement with experiments, which cover fluids with a range of surface tension values.
 
 Since the conclusions of \cite{bhagat2018origin} are markedly different from the accepted view of the role of gravity in these thin-film jumps the paper has attracted criticism, most particularly in a recent paper by \cite{bohr2019wrong} who, despite the experimental evidence, continue to question the role of surface tension in these flows. The underlying issue appears to be the current understanding of the interfacial flow which is that in both in hydrostatics and in hydrodynamics the influence of surface tension (when surface tension is uniform) is fully contained in interfacial pressure, commonly known as the Laplace pressure. The argument is that the force due to surface tension, which acts a tensile force at the line boundary of the interfacial surface, is mathematically equivalent to a pressure normal to the interface and, since the liquid velocity is tangential to the interface, this force can do no work. For hydrostatics, we have no argument with this viewpoint, {but in hydrodynamics, we will show that the interfacial surface energy is not conserved, except when the liquid film is completely flat and surface tension force is trivially balanced.} 
 
 
 Before proceeding further with this discussion we note that the importance of surface tension had previously been noted by \cite{mohajer2015circular} who  measured the jump radius and the height of the liquid downstream of the  jump for water and a water-surfactant solution (see {their} figures 4 and 13). They observed a significant increase in the jump radius when the surface tension was reduced {by the addition of the surfactant}. They also reported that for a range of flow rates the jump height remained constant and depended only on the surface tension of the liquid. 
 
  Despite this experimental evidence {several other recent papers } \citep{ wang2019role, PhysRevFluids.4.014002,fernandez2019origin,scheichlcentred2019} { continue to }support the previous gravity-based theory. For example, \cite{fernandez2019origin} performed numerical simulations {of the {\it downward}} vertical impingement of a liquid jet onto a horizontal plate and compared their results with the experimental data from \cite{button2010water}, {which were obtained for an {\it upward} directed jet onto a ceiling.}
   {Despite the opposite vertical orientations the authors concluded gravity is the dominant force.} 

{The role of surface tension is subtle, and one objective of this paper is to clarify the contribution of surface tension to the dynamics. }{The other objective is to show that the jump radius can be predicted using momentum conservation and  }show that {this is consistent with} the energy-based approach used by \cite{bhagat2018origin}. 
 {In order to focus on the role of surface tension} we predict the jump location using  conservation of momentum, ignoring gravity completely. 
 
 {Conventionally the circular jump has been} studied in an experimental arrangement where a vertical jet impinges onto a horizontal plate {and either flows over} a weir or off the edge of the plate.  In this paper we are concerned with the situation in which the plate is effectively an infinite plane {and we consider the jump} before the spreading liquid film encounters either the weir or the edge. The observations of \cite{bhagat2018origin} show \rkb{(for sufficiently high Reynolds numbers)} that the jump, once formed, has a constant radius until the spreading film reaches the edge of the plate. In practice, once the film reaches the edge of the plate a different boundary condition is imposed that changes the depth of the {downstream} subcritical film, which, in turn, changes the position of the previously-formed jump. Usually, in this later adjustment, since the subcritical region is signficantly deeper than the supercritical region upstream of the jump,  gravity can play a significant role. However, we will show that until the further downstream condition influences the jump, the initial jump location is determined by surface tension.

 The paper is organised as follows. We begin in \S\ref{sec:DA}{ by a consideration of dimensional analysis which will set the parameter ranges that are appropriate for our theoretical analysis.  The role of surface tension at the liquid-gas interface in a flowing, as distinct from stationary, liquid is derived and applied to a radially spreading thin film in \S\ref{sec:theory}.  A prediction for the jump radius is then obtained from application of conservation of momentum in both the radial and film-normal directions in \S\ref{sec:MC}. The energy-based approach presented in \cite{bhagat2018origin} is revisited and compared with the momentum based approach in \S\ref{sec:Relation to energy conservation}. {In \S\ref{sec:Relation to energy conservation1}, we discuss the relative importance of surface tension and re-visit the analysis of \cite{bohr2019wrong} and discuss the flaw in their theory.}    
Finally, our conclusions are given in \S\ref{sec:conc}.}
 \begin{figure}
\centering
\includegraphics[width=0.5\linewidth]{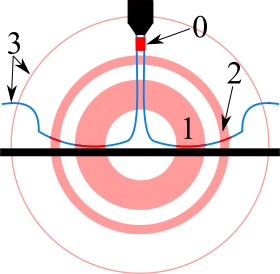}
\caption{{An illustrative sketch of a circular hydraulic jump on an infinite plate that was reported experimentally by \cite{bhagat2018origin}, along with a history of the interfacial surface area of a material fluid volume, $V = \pi a^2 l$ (indicated as $0$).} The annular surfaces are the top views of the surface of the material volume at different instances when it passes through locations $1$ and $2$ in the supercritical region and $3$ in subcritical region. The surface area of the  material volume is inversely proportional to the liquid film thickness $h$, which in the supercritical region where $h$ decreases implies an increase in surface area. At the location of hydraulic jump an abrupt change of liquid film thickness implies a release of the surface energy.}
\label{fig:Jump_area}
\end{figure}

\begin{table}
\centering 
\begin{tabular}[t]{c c c c c c c} 
 \vspace*{-1mm} 
Liquid & Reference &  $Q$ & $Q_c$ &$ \gamma$ $\times 10^{-3}$   & $ \nu$ $\times 10^{-6}$ & $\rho$ \\
\\

& & ($\SI{}{\centi\meter\cubed\per\second}$) & ($\SI{}{\centi\meter\cubed\per\second}$)& ($\SI{}{\kilo\gram\per\second\squared}$) &($\SI{}{\meter\per\second\squared}$)
&($\SI{}{\kilo\gram\per\meter\cubed}$)\\[0.5ex] 
\hline 
Water & \cite{mohajer2015circular} & 2.5-8.33 & 518& 72 & 1.002& 1000\\

Ethylene glycol & \cite{rojas2010inertial} & 20 &22& 45 & 7.6&1100\\

Silicon oil - 1  &\cite{duchesne2014constant} & 4.3-60 &2.3& 20 & 20&950-965\\ 

Silicon oil - 2  &\cite{duchesne2014constant} & - &0.45& 20 & 98&950-965\\

WP95/5 & \cite{bhagat2018origin} & 83-200 & 82 & 42.5 & 1.274&989 \\

WP80/20 & \cite{bhagat2018origin} & 83-200 & 58 & 26 & 2.30&968\\ 

SDBS & \cite{bhagat2018origin} & 83-200 & 147 & 38 & 1 &1000\\ 

\hline 
\end{tabular}

\caption{Parameters used in published experiments. The jet flow rate $Q$, the critical flow rate $Q_C$ at which gravity begins to play a role, the surface tension $\gamma$, the kinematic viscosity $\nu$, and the density $\rho$ of the fluid.\\ }
\label{tab:properties}
\end{table}

\section{Dimensional analysis}
\label{sec:DA}

We {begin by} considering the implications of dimensional analysis. The relative importance of gravity $g$ and surface tension depends on the fluid properties, {the value of the surface tension $\gamma$}, the density $\rho$ and kinematic viscosity $\nu$, and the flow characterised by the jet flow rate $Q$. Dimensional analysis shows that when {either }gravity or surface tension is \emph{ignored} the jump radius scales, respectively, as
\begin{eqnarray} \label{eq:R_ST}
R_{\textrm{ST}}  \sim  \frac{Q^{3/4}\rho^{1/4}}{\nu^{1/4}\gamma^{1/4}} \, \, {\textrm {or}} \, \, R_{\textrm{G}}  \sim  \frac{Q^{5/8}}{\nu^{3/8}{g^{1/8}}}.
\end{eqnarray}
{In a particular flow} the jump will either be caused by surface tension if $R_{\textrm{ST}} < R_{\textrm{G}}$ and and gravity if $R_{\textrm{ST}} > R_{\textrm{G}}$ .   {This criterion leads to a critical flow rate $Q_\textrm{C}$ given by
\begin{eqnarray}
Q_C = \frac{\gamma^2}{\nu \rho^2 g},
\end{eqnarray}
below which surface tension is the dominant force and above which gravity is important.  For water $Q_C \approx \SI{500}{\centi\meter\cubed\per\second}$ (i.e. $\approx 30$ L min$^{-1}$), which is significantly {larger}  than the flow in the standard kitchen sink.}

In the Appendix A (see (\ref{eq:SC1a})) we further show that including gravity provides a correction to the radius predicted by surface tension.
This correction to the pure surface tension radius $R_{\textrm{ST}}$
{takes the values} 0.95 and 0.77, for $Q = 2Q_C$ and $10Q_C$, respectively, which  shows that even for $Q \sim 10Q_C$,  the jump is dominated by the surface tension of the liquid.   Table~1 gives values of $Q_C$ for liquids used in experiments \citep{bohr1993shallow, bhagat2018origin, bohr2019wrong} and we see that $Q \ll Q_C$ in the experiments in water, and that the flow rates are only comparable in the experiments with ethylene glycol and silicone oil.  {However, as we note from (\ref{eq:SC1a}), even in these latter cases this scaling analysis implies that surface tension is still dominant. Consequently, we conclude that all the experiments listed in table \ref{tab:properties} are expected to fall within the range of parameters to which the theory developed below which ignores gravity is expected to apply.}

It is also worth noting that apart from surface tension the only parameters that have been varied in experiments are the viscosity and jet flow rate. The scaling relations (\ref{eq:R_ST}) have very similar dependences on these two parameters, which makes it very hard to distinguish between them on experimental grounds alone. This is possibly why some controversy persists about the interpretation of the experimental observations.

 \section{{Theory}}
\setlength{\belowcaptionskip}{-10pt} 
 \label{sec:theory}
 \subsection{Surface forces and energy in hydrodynamics}
 \label{subsec:Surface energy in hydrodynamics}

  Surface forces {at a liquid interface} arise from short range intermolecular interactions and the total short range force acting on a {fluid} element is determined by the surface area of the element, {while} the volume of the element is not relevant for such forces \citep{batchelor2000introduction}. {The molecular origin of surface tension and surface energy is associated with intermolecular cohesive forces. The average free energy associated with a fluid molecule in the liquid bulk, where it is surrounded by similar molecules is independent of its position. However, within a distance less than the range of the cohesive forces ($ 10^{-9}\SI{}{\meter})$ from the the interface, the liquid molecules have an additional free energy which is proportional to the interfacial surface area.} A net exchange of liquid molecules between the interface and the bulk that necessarily accompanies a change in surface area will be associated with a net energy exchange. 
  
  { Now consider the case of an impacting liquid jet and  follow the small material volume of fluid, $\pi a^2 l$ in the jet (indicated as $0$ in figure \ref{fig:Jump_area}), where $a$ is the radius of the jet and $l$ is the length of the cylindrical fluid element. The surface area of a liquid interface changes as it spreads on the plane as a thin film of thickness $h$, and it is straightforward to show that the ratio of the final to initial surface area is $a/2h$. In the spreading film when $h$ decreases with radius and $a/2h \gg 1$, implying that there is an energy penalty as liquid molecules leave the bulk to increase the surface area of the spreading film. Whereas, at the hydraulic jump, an abrupt increase of liquid film thickness implies a release of the surface energy. This clearly demonstrates that even in  a steady flow a spatial variation of liquid film thickness implies an active exchange of mechanical and surface energy. } We now discuss the way in which this exchange of energy is accounted for.
  
  \subsection{Surface tension and the interfacial boundary condition }
 \label{ssec:STF}
 
 Fluid motion is governed by the Navier-Stokes equations which express  conservation of momentum in a fluid continuum. For flows with an interface between two fluids, the Navier-Stokes equations do not express the surface tension force acting on the interface. This force is introduced as a normal stress boundary condition at the interface.

Consider a fixed surface $S$, with unit normal $\nb$, bounded by a closed contour $C$ with the line element $d\textbf l$ around the contour in the interface between two immiscible fluids, taken here for simplicity to be the common case of a liquid and a gas denoted by the subscripts $L$ and $G$, respectively, with constant surface tension $\gamma$ (see figure~\ref{Hydraulic_jump_fig}). Since the surface tension force acts in a direction perpendicular to $\nb$  and the contour $C$, continuity of the normal stress is expressed as 
 \begin{eqnarray}
 {\int\limits_S (\Tb_G - \Tb_L)\cdot \nb dS + \gamma\int\limits_C   d{\textbf l} \times \nb  =  0,}
     \label{eq:stress}
 \end{eqnarray}
 where $\Tb = -p\textbf{I} + \mu [\nabla \ub + (\nabla \ub)^T] $ is the total stress, with pressure $p$ and velocity $\ub$, and $\mu$ is the {dynamic} viscosity of the fluid. 
 Using the vector identity and {$\int \limits_C d\textbf{l} \times \nb  \rkb{\equiv} -\int \limits_S  (\nabla_s \cdot \nb) \nb dS $},
 \rkb{
 \begin{eqnarray} \label{eq:normal_stress0}
 \int\limits_S(\Tb_G - \Tb_L) \cdot \nb dS & = &    \int\limits_S\gamma (\nabla_s \cdot \nb) \nb dS.
 \end{eqnarray}
 } \PFL{Since the surface $S$ is arbitrary }we then obtain the conventional form of the dynamic boundary condition 
  \begin{eqnarray} \label{eq:normal_stress}
 (\Tb_G - \Tb_L) \cdot \nb & = &   \gamma (\nabla_s \cdot \nb) \nb,
 \end{eqnarray}
where $\nabla_s = [\textbf{I} - \textbf{nn}] \cdot\nabla$ is the surface gradient, relating the jump in the  normal stress to the curvature of the surface in agreement with \cite{bush2003influence} \PFL{and valid pointwise in space}.  
\rkb{However, \PFL{it is important to note that} the surface tension forces act along the local tangents at different points of the closed contour $C$, and \PFL{it is} the net resultant of these tensile forces, acts along the normal to the interface.} 
 In the case of a stationary liquid and gas, the viscous term is zero, and this equation gives the usual Laplace pressure in the liquid associated with the curvature of the surface. In the case of a flowing liquid \rkb{ (\ref{eq:normal_stress}) will give the correct net resultant force on an interfacial surface $S$, \PFL{but does not} give a detailed account of the tensile forces \PFL{along} the contour, $C$.}  The  velocity of  the liquid may vary around  the contour $C$ implying that liquid molecules crossing the boundary may not be conserved and need to be accounted for in the surface energy exchange. We consider this effect in the next section.

\subsection{Interfacial surface energy conservation}
\label{ssec:isec}

 Assuming the \textit{dynamic} viscosity of the gas (for air and water the ratio  $\mu_a/\mu_w \sim 10^{-2}$) is negligible compared to that of the liquid \pfl{and that the velocities in the liquid and gas are comparable,} and denoting pressure in the gas as $p_G$  and in the liquid as $p_L$,  (\ref{eq:stress}) can be written as 
 
  \begin{equation}
     \int\limits_S (-p_G + p_L) \textbf{n} \, dS - \mu \int\limits_S  \textbf{n} \cdot[\nabla \ub + (\nabla \ub)^T]  dS + \gamma \int \limits_C  d{\textbf l} \times \nb = 0,
    \label{eq:nstress}
 \end{equation}
 where ${\bf u}$ is the velocity in the liquid \emph{at the interface}. 
 {The surface energy flux can be calculated by taking a dot product between the interfacial velocity at the contour $C$ and the forces to obtain (since $\ub \cdot \nb = 0$ on the interface)
}
\begin{equation}
      \mu \int\limits_S \textbf{u} \cdot \left(\textbf{n}.[\nabla \ub + (\nabla \ub)^T]\right)  dS = \gamma \int \limits_C  \textbf{u} \cdot (d{\textbf l} \times \nb),
    \label{eq:nstress_energy1}
 \end{equation}
which can be re-expressed as

  \begin{eqnarray} \label{eq:nstress_energy3}
        \mu \int\limits_S  \left\{\textbf{u} \cdot [(\textbf{n} \cdot \nabla) \ub] + \nb \cdot [(\textbf{u} \cdot\nabla) \ub] \right\}  dS &  = & \gamma \int \limits_S \{ (\nabla \cdot \ub)- [(\nb \cdot  \nabla) \ub] \cdot \nb\} dS,\\ & = & \gamma \frac{DS}{Dt}, \nonumber
 \end{eqnarray}
 the rate of change of the material surface area $S$ (see (3.1.5) of \cite{batchelor2000introduction}).

{Thus the increase in surface energy is supplied by the kinetic energy of the flow, which, of course, is zero in the case of stationary liquid. {We also emphasise} that  the velocity field here is the surface velocity field, for which $\nabla \cdot\ub$ may or may not be equal to zero {due to the addition (or loss) of fluid elements to the liquid interface as discussed above}.} 
 The surface energy will only be conserved when velocity field at the interface satisfies the condition $[( \nabla \cdot\ub)- [(\ub \cdot \nabla)\nb] \cdot \nb  = 0$, (which is satisfied for a flat surface {for which net force due to surface tension is zero})  otherwise there will be an active energy exchange with the bulk. 

\PFL{It is important to re-emphasise} \rkb{ that the pointwise dynamic boundary condition in its conventional form  (\ref{eq:normal_stress}), which if used to calculate the surface energy \PFL{over a surface $S$} does not account for the variation in velocity along the contour $C$,  will give a flawed result  \PFL{implying that} the net surface energy exchange \PFL{is} zero.} \PFL{We emphasise that (\ref{eq:nstress_energy3}) is the \textit{correct interfacial boundary condition} to be applied at the liquid interface and will be used in the analysis below.}

 \section{Momentum conservation}
\label{sec:MC}

We now apply conservation of momentum to the axisymmetric flow spreading radially from the point of impact of the jet on the plane, ignoring gravity. {In cylindrical coordinates ($r, \theta)$ with the origin at the point of jet impact, let $u, w$ be the  radial and vertical velocity components, respectively.} 
 
 First we determine the  surface boundary condition.  Following \cite{bush2003influence}
 we write the equation of the axisymmentric surface {$h(r)$} in implicit form  
 \begin{equation}
     J(r,z) = z - h(r) = 0,
     \label{surface}
 \end{equation}
which yields the  vector {normal to the film surface} 
 \begin{equation}
     \textbf{n} = \frac{\nabla J}{\abs{\nabla J}} = \frac{\Hat{\zb} - h' \Hat{\rb}}{(1+h'^2)^{1/2}},
      \label{surface_normal}
 \end{equation}
 where $\hat{\rb}$ and $\hat{\zb}$ are unit vectors in the radial and wall-normal directions, respectively, and $h' = {dh}/{dr}$.
 We define the angle $\alpha$ as the tangent to the surface defined by $h' = \tan\alpha$. Then $\cos\alpha = \frac{1}{(1 +  (h'^2)^{1/2}}$ and $\sin\alpha = \frac{h'}{(1 +  h'^2)^{1/2}}$, and (\ref{surface_normal}) can also be written as 
 \begin{equation}
     \textbf{n} = \hat{\zb}\cos\alpha  - \hat{\rb}\sin\alpha.
     \label{norm2}
 \end{equation}
We denote the radial velocity component at the surface $z = h$ as $u_s(r)$. From the kinematic boundary condition the vector surface velocity can be written as  
\begin{equation}
     \ub = u_s (r) \hat{\rb} + u_s (r) h^\prime (r) \hat{\zb},
     \label{vel2}
 \end{equation}
  which implies that, at the interface

\begin{equation}
       [(\nabla \cdot\ub)- [(\nb \cdot \nabla) \ub] \cdot \nb    =  \frac{1}{r}\frac{d (u_s r)}{dr} + \frac{u_s h^\prime h^{\prime \prime} }{(1+h^{\prime^2})} = \frac{1}{r(1+ {h^\prime}^ 2)^{(1/2)}}\frac{d\{u_sr (1+ {h^\prime}^ 2)^{(1/2)}\}}{dr}.
    \label{eq:boundarycond}
 \end{equation}
  Substituting (\ref{eq:boundarycond}) in (\ref{eq:nstress_energy3}) and recognising \pfl{ $dS = r(1+ {h^\prime}^ 2)^{(1/2)} d\theta dr $} yields,
  \begin{eqnarray} \label{eq:final}
        \mu \int\limits_S  \left\{\textbf{u} \cdot [(\textbf{n} \cdot \nabla) \ub] + \nb \cdot [(\textbf{u} \cdot\nabla) \ub] \right\}  dS &  = & \gamma \int \limits_S\frac{d\{u_sr (1+ {h^\prime}^ 2)^{(1/2)}\}}{dr} dr d\theta,
 \end{eqnarray}
  
 Note that these results depend only on the surface velocity and holds independent of the wall-normal velocity profile.

However, in order to obtain an explicit expressions we follow \cite{watson1964radial} and write the velocity in similarity form

 \begin{equation} \label{eq:u}
 u(r, z) = u_s(r) f(\eta), \; \; \eta \equiv \frac{z}{h(r)}, \; 0 \leq \eta \leq 1,
 \end{equation}
 where $u_s$ as above is the surface velocity and $f(0) = 0, f(1) = 1$. Conservation of mass implies
 
 \begin{equation} \label{eq:mass}
     \int_0^h ur dz =  r u_s h \int_0^1 f(\eta)d\eta \equiv C_1 u_s r h = Q/2\pi,
 \end{equation}
 and $C_1 \equiv \int_0^1 f(\eta)d\eta$ is an integration constant arising from the  velocity profile. Incompressibility gives
 
 \begin{eqnarray}
 w  =  - \int_0^{\eta} \frac{1}{r} \frac{\partial {ru}}{\partial {r}} dz
  =  - \frac{1}{r} \frac{d }{dr} [ (r u_s h) \int_0^{\eta} f(\eta) d\eta]. \nonumber
 \end{eqnarray}
 Then, using (\ref{eq:mass}) 
 
 \begin{eqnarray} \label{eq:w}
 w  =  - u_s h \frac{d}{dr} \int_0^{\eta} f(\eta) d\eta
  =  u h' \eta = u_s h' \eta f(\eta),
 \label{equation3.4}
 \end{eqnarray}
 {which automatically satisfies the kinematic boundary conditions at the wall and interface.}  We now use these expressions for the velocity in equations expressing momentum conservation in the radial and wall-normal directions.

 \begin{figure}
\centering
\includegraphics[width=0.8\linewidth]{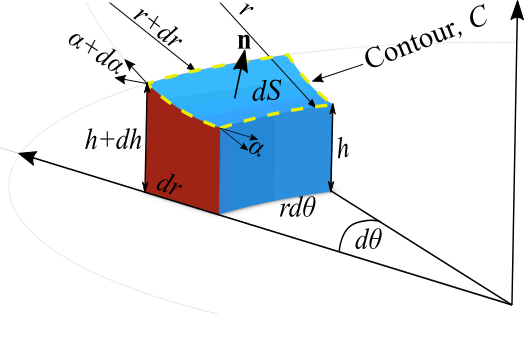}
\label{fig:HJ_pic}

\setlength{\belowcaptionskip}{-10pt}
\vspace{5pt}

\caption{{ A schematic of the differential control volume showing the slope of the thin liquid film, and a contour, $C$ enclosing an interfacial surface, $S$.  } }
\label{Hydraulic_jump_fig}
\end{figure}

\subsection{Force on an axisymmetric thin film}
\label{ssec:thin_film}

{For the control volume shown in figure~\ref{Hydraulic_jump_fig}, with the interface bounded by the closed contour $C$, the differential arc length $d{\textbf l} = \left( r  d\theta \right)\hat{{\boldsymbol\theta}} \Bigr|_{\substack{r}}^{\substack{r+ dr}} - \left( dr \right)\hat{\rb} \Bigr|_{\substack{\theta}}^{\substack{\theta+  d\theta}} -\left( dh \right)\hat{\zb} \Bigr|_{\substack{\theta}}^{\substack{\theta+  d\theta}}. $ Consequently, the force due to surface tension on this interface is

\begin{eqnarray} 
\begin{split}
dF_\gamma = \gamma \left(d{\textbf l}\times \textbf{n}\right) 
 =  \gamma\left(d\theta r \cos\alpha \hat{\rb}\right)\Bigr|_{\substack{r}}^{\substack{r+ dr}} + \gamma\left( d\theta r \sin\alpha \hat{\zb}\right) \Bigr|_{\substack{r}}^{\substack{r+ dr}} + \gamma\left(\frac{dr}{\cos\alpha}  \hat{\theta} \right) \Bigr|_{\substack{\theta}}^{\substack{\theta+ d\theta}}.
\end{split}
\label{eq:ST1}
\end{eqnarray}
Considering the circular symmetry, in the limit of $dr $ and $ d\theta \to 0 $ 
 \begin{equation}
     dF_\gamma = \left(\gamma {d (r\cos\alpha)}  - \gamma \frac{dr}{\cos\alpha }\right)d\theta \hat{\rb} + \gamma {d (r\sin\alpha)}  d\theta\hat{\zb}.
     \label{eq:STForce}
 \end{equation}
 
Therefore, from (\ref{eq:nstress_energy1}) and (\ref{eq:STForce}), the radial and vertical components of the normal stress at the free surface can be written as }

\begin{equation}
\begin{split}
 F_{\gamma, r} \equiv \int_r^{r+dr} \mu\left(  \textbf{n} \cdot [\nabla \ub + (\nabla \ub)^T]\right) \cdot \hat{\rb} \left( r d \theta \frac{d r}{\cos\alpha} \right)  \Bigr|_{\substack{h}} = \int_r^{r+dr} P \sin\alpha rd \theta\frac{d r}{\cos\alpha} \\ +  \gamma r d\theta \cos\alpha\Bigr|_{\substack{r}}^{\substack{r+ dr}} - \gamma \frac{dr}{\cos \alpha}d \theta,
 \end{split}
  \label{eq:RS}
 \end{equation} 
 and
 \begin{equation}
 F_{\gamma, z} \equiv \int_r^{r+dr} \mu\left(  \textbf{n} \cdot [\nabla \ub + (\nabla \ub)^T]\right) \cdot \hat{\zb} \left( rd \theta\frac{d r}{\cos\alpha} \right)  \Bigr|_{\substack{h}} = \int_r^{r+dr} P rd \theta d r + rd \theta \gamma \sin\alpha\Bigr|_{\substack{r}}^{\substack{r+ d r}},
  \label{equation_vertical_stress}
 \end{equation}   
 where $P = p_L - p_G$. {As we will see below, application of momentum conservation requires expressions for the radial gradients of these forces, and as axisymmetry implies we can drop $d\theta$, these are given by 

\vspace{-10pt}

\begin{equation}
\begin{split}
  \frac{d F_{\gamma, r} }{dr} = \frac{d}{dr}\left(\int_r^{r+dr} \mu\left(  \textbf{n} \cdot [\nabla \ub + (\nabla \ub)^T]\right)\cdot \hat{\rb} \left(\frac{ r dr}{\cos\alpha} \right)  \Bigr|_{\substack{h}}\right)  =   r P \tan\alpha -\frac{\gamma}{\cos \alpha} + \gamma \frac{ d(r  \cos\alpha)}{dr}\\
  \label{equation_radial_stress}
  \end{split}
 \end{equation} 
and

\vspace{-10pt}
 \begin{equation}
   \frac{d F_{\gamma, z} }{dr}      =  \frac{d}{dr}  \left(\int_r^{r+dr}\mu\left(  \textbf{n} \cdot [\nabla \ub + (\nabla \ub)^T]\right)\cdot\hat{\zb} \left(\frac{ r dr}{\cos\alpha} \right)  \Bigr|_{\substack{h}}\right) =   r P   +  \gamma \frac{d(    r\sin\alpha)}{dr}.
  \label{equation1_z_compo}
 \end{equation} 
}

 \vspace{-10pt}

\subsection{Momentum balance in the radial direction}
\label{Momentum balance in radial direction}

 In the absence of gravity, 
 the momentum equation in the radial direction is 
 \begin{equation}
     \rho \left( u  \frac{\partial u}{\partial r} + w  \frac{\partial u}{\partial z}  \right) =  \frac{1}{r} \frac{\partial (r\tau_{rr})}{\partial r} + \frac{\partial \tau_{rz} }{\partial z}, 
     \label{nav_sto_eq}
 \end{equation}
where $\tau = \TT + p\textbf{I}$ is {the deviatoric stress tensor $\TT$ and pressure} (in cylindrical coordinates). Since the pressure in the gas at the surface is constant (atmospheric) and the film is thin, the radial pressure gradient in the liquid is only a function of radius caused by the curvature of the surface.

We use incompressibility ($\nabla \cdot \ub = 0$ { in the bulk fluid}), integrate (\ref{nav_sto_eq}) across the film from the wall to the interface, use axisymmetry to eliminate $d\theta$, substitute for the velocity from (\ref{eq:u}) and (\ref{eq:w}) and apply the surface tension boundary condition (\ref{eq:boundarycond}),  to obtain
\begin{equation}
\begin{split}
\frac{d}{dr}\int_0^h \rho r u^2 dz 
 = -\int_0^h \frac{d p}{d r}r dz + r p \tan\alpha - \frac{\gamma}{\cos\alpha} +  \gamma \frac{ d(r  \cos\alpha)}{dr}  -\mu r \left(\frac{\partial {u}}{\partial {z}} + \frac{\partial {w}}{\partial {r}} \right) \Bigr|_{\substack{0}}.
\label{momentum_balance_interstep_1}
\end{split}
\end{equation}
Noting that conservation of mass (\ref{eq:mass}) implies  $u_srh = \textrm{const.} $    we obtain

\begin{equation}
\begin{split}
C_2[\rho u_s r h \frac{  d {u_s} }{d r} ]
 = -\frac{d p}{d r} rh + r p \tan\alpha - \frac{\gamma}{\cos\alpha} +  \gamma \frac{ d(r  \cos\alpha)}{dr}   -\mu r \left(\frac{\partial {u}}{\partial {z}} + \frac{\partial {w}}{\partial {r}} \right) \Bigr|_{\substack{0}},
\label{momentum_balance_equation_Bush2}
\end{split}
\end{equation}
where $C_2 = \int_0^1 f^2 (\eta) d \eta$ is a second integration constant.
Equation (\ref{momentum_balance_equation_Bush2}) can be written in terms of the interface slope
\begin{eqnarray}
\label{momentum_balance_equation17}
C_2[\rho u_s r h \frac{  d {u_s} }{d r} ] 
 & = & -\frac{d p}{d r} rh +  r p h' - {\gamma}{(1+ h'^2)^{1/2}} + \gamma \frac{1}{(1+ h'^2)^{1/2}}\\ \nonumber
& & - \gamma\frac{r h'h''}{(1+ h'^2)^{3/2}} - \tau_w r,
\end{eqnarray}
where $\tau_w = -\mu \left(\frac{\partial {u}}{\partial {z}} + \frac{\partial {w}}{\partial {r}} \right) \Bigr|_{\substack{0}}$ is the wall shear stress.

\subsection{Momentum balance in wall-normal direction}
\label{Momentum balance in wall-normal direction}
We now apply conservation of momentum in the wall-normal $z$ direction in the differential control volume shown in figure \ref{Hydraulic_jump_fig}. Since the film is thin the pressure is independent of $z$ and for an axisymmetric flow

\begin{equation}
    \rho u r  \frac{\partial w}{\partial r} + \rho w r \frac{\partial w}{\partial z}  = \frac{\partial (r\tau_{rz}) }{\partial r} + r\frac{\partial \tau_{zz}}{\partial z}.
    \label{NV_Zdirection}
\end{equation}
As before we integrate across the film  to obtain
\begin{equation}
   \int_0^h \rho u r \frac{\partial w}{\partial r} dz + \int_0^h \rho w r \frac{\partial w}{\partial z} dz = \int_0^h \frac{\partial (r\tau_{rz}) }{\partial r}dz \rkb{{- 2 \mu r \frac{\partial w}{\partial z}|_h}}.  
    \label{NV_Zdirection_int_1}
\end{equation}
Substituting for the velocity from (\ref{eq:u}) and (\ref{eq:w}) gives

\begin{equation}
\begin{split}
  \rho u_s r h \frac{d(u_s h^\prime)}{dr}\int_0^1 \eta f^2(\eta) d\eta - \rho h^{\prime 2} u_s^2 r \int_0^1 \eta f^2(\eta) d\eta - \rho \frac{u_s^2 r h^{\prime 2}\eta^2 f^2(\eta)}{2}  \Bigr|_{\substack{0} }^{1} \\+ \rho h^{\prime 2} u_s^2 r \int_0^1 \eta f^2(\eta) d\eta + r\frac{\rho w^2}{2}  \Bigr|_{\substack{0} }^{h} = \frac{d F_{\gamma, z}}{dr} -\int \frac{\partial (r\tau_{rz}) }{\partial r}dz \Bigr|_{\substack{0} } \rkb{{- 2 \mu r \frac{\partial w}{\partial z}|_h}}.
  \end{split}
  \label{euuation 20}
\end{equation}
Applying the surface boundary condition (\ref{equation1_z_compo}) into (\ref{euuation 20}) yields 

\begin{eqnarray}
 \label{NV_Zdirection_abcd}
\rho u_s r h \frac{d (u_s h^\prime)}{d r} \int_0^1 \eta f^2(\eta) d\eta &   = & {r P}  +  \gamma \frac{d(    r\sin\alpha)}{dr} - \int \frac{\partial (r\tau_{rz}) }{\partial r}dz \Bigr|_{\substack{0} } \rkb{{- 2 \mu r \frac{\partial w}{\partial z}|_h}},   \\
&  = & rP+ \gamma  \sin\alpha + \gamma r \cos\alpha\frac{d \alpha}{dr}- \int \frac{\partial (r\tau_{rz}) }{\partial r}dz \Bigr|_{\substack{0} } \rkb{{- 2 \mu r \frac{\partial w}{\partial z}|_h}}, \nonumber \\
& = & rP   +   \frac{\gamma h'}{(1+ h'^2)^{1/2}} + \frac{\gamma r h''}{(1+ h'^{3/2})}- \int \frac{\partial (r\tau_{rz}) }{\partial r}dz \Bigr|_{\substack{0} } \rkb{{- 2 \mu r \frac{\partial w}{\partial z}|_h}}. \nonumber
 \end{eqnarray}

 In the thin liquid film upstream of the hydraulic jump, the interface slope remains small and  we will ignore the higher order terms in $\frac{dh}{dr}$.  Applying this approximation and re-arranging (\ref{NV_Zdirection_abcd}) gives an expression for the curvature of the film
 \begin{equation}
 \begin{split}
h''(\rho u_s^2 r h  \int_0^1 \eta f^2(\eta) d\eta - \gamma r) + h' (\rho u_s r h\frac{du_s}{dr}\int_0^1 \eta f^2(\eta) d\eta  - \gamma) = \\r P   - \int \frac{\partial (r\tau_{rz}) }{\partial r}dz \Bigr|_{\substack{0} } \rkb{{- 2 \mu r \frac{\partial w}{\partial z}|_h}}.
\label{NV_Zdirection_int3}
\end{split}
 \end{equation}
Finally, substituting $\rho u_s r h \frac{du_s}{dr}$ from the radial momentum balance  (\ref{momentum_balance_equation17}) gives 
 \begin{equation}
 \begin{split}
[C_3\rho u_s^2 rh -\gamma r] h'' +  \frac{C_3}{C_2} h' ( -\frac{d P}{d r} rh +  r P h' - {\gamma}{(1+ h'^2)^{1/2}} + \gamma \frac{1}{(1+ h'^2)^{1/2}} - \\ \gamma\frac{h'h''}{(1+ h'^2)^{3/2}} 
 - \tau_w r - \frac{C_2}{C_3}\gamma ) 
 =  r P   - \int \frac{\partial (r\tau_{rz}) }{\partial r}dz \Bigr|_{\substack{0} } \rkb{{- 2 \mu r \frac{\partial w}{\partial z}|_h}},
\end{split}
\label{NV_Zdirection_int_4}
 \end{equation}
 where $C_3 = \int_0^1 \eta f^2(\eta) d\eta$ {is a third integration constant}.

 Ignoring higher order terms in $h'$ and re-arranging  gives an expression for the curvature of the film
 
 \begin{equation}
h'' =  \frac{-h'\frac{C_3}{C_2}[-\frac{d P}{d r}rh + r\tau_w - \frac{C_2}{C_3}\gamma] + rP + \int \frac{\partial (r\tau_{rz}) }{\partial r}dz \Bigr|_{\substack{0}} \rkb{{- 2 \mu r \frac{\partial w}{\partial z}|_h}}}{C_3\rho u_s^2 r h (1 - {\textrm We}^{-1}) } 
 \end{equation}
 where the Weber number $We$ is defined by
 
 \begin{equation}
     {\textrm We} \equiv \frac{C_3 \rho u_s^2 h}{\gamma}.
 \end{equation}
 Consequently, we predict a singularity in the curvature of the film at a critical radius where the film thickness is such that $We = 1$. This criterion gives the location of the hydraulic jump.
 
\begin{figure}
\centering

\begin{subfigure}[h]{0.4\linewidth}
\includegraphics[width=\linewidth]{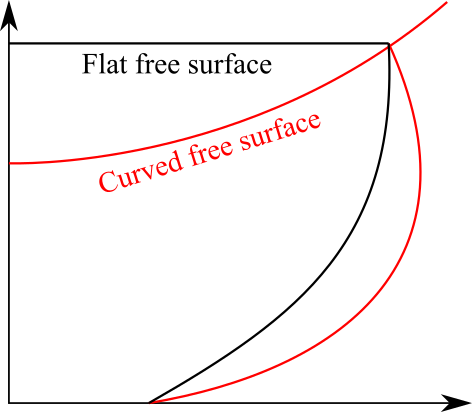}
\caption{  Velocity profile}
\end{subfigure}%
\hspace{8 mm}
\begin{subfigure}[h]{0.4\linewidth}
\includegraphics[width=\linewidth]{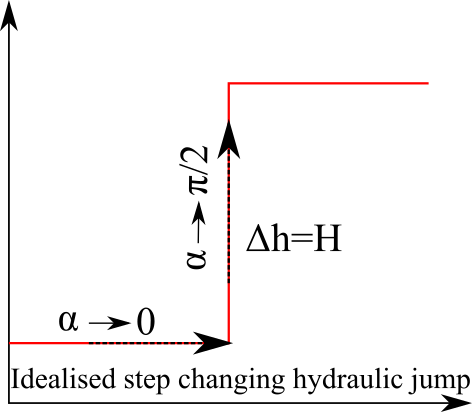}
\caption{An idealised jump }
\end{subfigure}
\setlength{\belowcaptionskip}{-10pt}
\vspace{5pt}
\caption{ (a) Schematic velocity profiles in a flow with   conventional flat surface assumption with zero radial viscous stress (black) and liquid interface with zero tangential stress but non-zero radial stress (red). The surface tension force retards the flow in the radial direction and accelerates it in the wall-normal direction, giving  non-zero radial and wall-normal viscous stresses at the surface (\ref{eq:stress}) (b) An idealised hydraulic jump. }
\label{fig:Vel_profile}
\end{figure}

 {\subsection{Revisiting the radial momentum balance}
 
 We now revisit  the radial momentum balance (\ref{momentum_balance_equation17}) which can also be written as  
 \begin{eqnarray}
C_2[\rho u_s r h   d {u_s} ] 
 & = & - rh dp +  r p dh - {\gamma}{(1+ h'^2)^{1/2}} dr + \gamma \frac{dr}{(1+ h'^2)^{1/2}}\\ \nonumber
& & - \gamma\frac{r h'h'' dr}{(1+ h'^2)^{3/2}} - \tau_w r dr,
\label{RADIALSCA}
\end{eqnarray}
 and apply it at the jump radius.
We note that the jump is a singularity (see figure \ref{fig:Vel_profile}(b)), where $\frac{dh}{dr}= \tan\alpha \to \infty$, $dr \to 0$, $dh = H $, a finite quantity, and $\alpha$ changes from $0 $ to $\pi/2$\rkb{, and the radial velocity changes from $u_s$ to $0$.} Substituting the trigonometric forms of the functions and integrating (3.15) at the jump location $r = R$ gives 
 \begin{eqnarray}
\int C_2\rho u_s R h   d {u_s}  
 & = & - \int Rh dp +  \int R p dh - \int {\gamma}\sin\alpha dh \rkb{+\gamma R \cos\alpha \Bigr|_{\substack{0}}^{\substack{\pi/2}}} \\ \nonumber
& & + \int \gamma \cos\alpha dr - \int \tau_w R dr.
\label{RADIALSCA1}
\end{eqnarray}
Then in scaled terms  (\ref{RADIALSCA}) can be written 
\rkb{
\begin{equation}
    -C_2\rho {u_s}^2 R h \approx RH (p- \frac{\gamma}{R}) - \gamma R.
\end{equation}
}
Since, at the jump, the pressure $p$ scales as $\frac{\gamma}{R}$ (see \cite{bush2003influence}), the first term on the right hand side of (\ref{RADIALSCA1}) is zero, which gives $We = 1$ as the condition for the hydraulic jump. Thus conservation of  radial momentum gives \pfl{a similar} result for the jump condition.}

 \section{Relation to energy conservation }
 
 \label{sec:Relation to energy conservation}
{ For a better physical understanding we can also get the interfacial energy from a consideration of the interfacial force. From (\ref{eq:STForce}) the energy flux associated with surface tension force on the control volume (figure~\ref{Hydraulic_jump_fig}) is}
 \begin{eqnarray} 
\begin{split}
dF_\gamma. (u \hat{\rb} + w \hat{\zb}) = \gamma\left(d\theta u r \cos\alpha \right)\Bigr|_{\substack{r}}^{\substack{r+ dr}} + \gamma\left( d\theta w r \sin\alpha \right) \Bigr|_{\substack{r}}^{\substack{r+ dr}}.
\end{split}
\label{Energyeq1}
\end{eqnarray}

Circular symmetry implies that there is no net flux of fluid in the azimuthal direction and so substituting (\ref{vel2}) into (\ref{Energyeq1}) yields  

  \begin{eqnarray} 
\begin{split}
dF_\gamma. (u \hat{\rb} + w \hat{\zb}) = \gamma\left( d\theta u_s r(1+h^{\prime 2})^{1/2}  \right) \Bigr|_{\substack{r}}^{\substack{r+ dr}},
\end{split}
\label{Energyeq3_2}
\end{eqnarray}
which is the equivalent result as (\ref{eq:final})  derived by considering rate of change of interfacial surface area and work done by the surface tension force.  Consequently, in a control volume approach the force due to surface tension, which  appears through the normal stress boundary condition, can be incorporated as a surface force on the circumference of the control volume  consistent with the analysis in \cite{bhagat2018origin}.

\section{Relative importance of surface tension}
 \label{sec:Relation to energy conservation1}
{We now return to the question of the the relative importance of gravity and surface tension in these flows}. \cite{bohr2019wrong}
{ argued} that the analysis of \cite{bhagat2018origin} is wrong and that surface tension does not play a significant role in the formation of these thin-film jumps. \cite{bohr2019wrong} wrote the {energy} equation over a volume $V$ bounded by a closed surface $A$ (their equation (13)) as
\begin{equation}
 -\int_A \left[v_j \left( \shalf \rho v_i v_i + p\right) - \mu v_i v_{i,j} \right] n_j dA - \shalf\mu \int_V v_{i,j}^2 dV = 0, 
 \label{Bohr1}
\end{equation}
where the pressure $p$ is taken to be sum of a Laplace term and a hydrostatic pressure term.
 \cite{bohr2019wrong} initially ignore all the viscous terms in (\ref{Bohr1}) and integrate the first term over a cylindrical surface spanning the film to obtain (in our notation)
 
 \begin{equation}
\chi(r) = \int_0^h u \left(\frac{1}{2}\rho (u^2+w^2) + p\right)r dz \approx   \shalf C_3 \rho r u_s^3 h  + C_1 u_s r h p.  
 \label{Bohr2_2}
\end{equation}
 They then differentiate this expression with respect to $r$, and set the derivative $\chi'$ equal to the viscous terms which they denote as $\xi$. This leads to their equation (20), where the viscous terms remain unspecified as $\xi$. However, the surface energy term  $\int \mu v_i v_{i,j}n_jdS$ evaluated at the \emph{free surface} $S$ is, from  (\ref{eq:nstress_energy3}), 
\begin{equation}
 \begin{split}
   \int_S \mu v_i v_{i,j}n_jdS 
   \equiv \mu \int\limits_S  \left\{\textbf{u}.[(\textbf{n}.\nabla) \ub] + \nb. [(\textbf{u}.\nabla) \ub] \right\}  dS  =  \gamma \int \limits_S \{ (\nabla.\ub)- [(\nb. \nabla) \ub].\nb\} dS, 
 \end{split}
    \label{eq:nstress_energy_bohr}
 \end{equation}
and includes the effect of surface tension and depends on the surface velocity. \pfl{In their analysis, \cite{bohr2019wrong} have ignored the surface energy term.}

\pfl{\cite{bohr2019wrong}integrated (\ref{Bohr1}) over the inner cylinder of the control volume}. Instead, in our analysis we integrate (\ref{Bohr1}) for the complete closed annular control volume from 0 to $2\pi$ which yields, 
\begin{equation}
    - \shalf C_3 \rho r u_s^3 h  \Bigr|_{\substack{r}}^{\substack{r+ dr}} - C_1 u_s r h p  \Bigr|_{\substack{r}}^{\substack{r+ dr}} + \gamma {d\{u_sr (1+ {h^\prime}^ 2)^{(1/2)}\}}  - \tau_w r dr = 0.
    \label{Bohr_integral}
\end{equation}
 Recognising that $u_srh =$ constant, $\tau_w$ is the wall friction and dividing (\ref{Bohr_integral}) by $dr$, the limit of $h^\prime \to 0$ gives, 
 \begin{equation}
    - C_3 \rho r u_s^2 h \frac{d u_s}{dr} + \gamma r  \frac{d u_s }{dr} + \gamma u_s - C_1 u_s r h \frac{d p}{dr} +  \tau_w r dr = 0
    \label{Bohr_integral_2}
\end{equation}
This equation is equivalent to (in the absence of gravity) (5.4) in \cite{bhagat2018origin} which was obtained from a control volume approach and leads to a critical Weber number of order one at the jump (and not $O(\alpha)$ as \cite{bohr2019wrong} claim).

Including this surface term  explicitly (\ref{eq:nstress_energy_bohr}) in $\xi$, since it involves $u_s$ and $\gamma$ will change the result in \cite{bohr2019wrong} and lead to the conclusion that surface tension is the dominant force, consistent with the increase in surface area documented in \S\ref{ssec:isec}.

\section{{Conclusions}}
\label{sec:conc}
Applying conservation of radial and wall-normal momentum to the flow in an expanding axisymmmetric thin film shows that the curvature of the film is singular at a finite radius determined by a critical value of the Weber number. This singularity arises from the wall-normal momentum conservation which implies that $\frac{d^2h}{dr^2}\to \infty$ whereas $(\frac{dh}{dr})^2 \to 0$ at a finite radius. In \S3.1, we presented a physical mechanism to demonstrate the mechanical and surface energy exchange in steady flow. Physically, in the thin film limit, $ h^{\prime \prime} = (1+h^{\prime 2})\frac{d \alpha }{dr} \approx  \frac{d \alpha }{dr} $, and the singularity implies that this change in interface slope $\frac{d \alpha }{dr}$ coincides with the hydraulic jump. Therefore, the mathematical expression in tandem with the physical interpretation indicates a jump. We also show that \rkb{a similar} jump condition can be obtained by applying conservation of radial momentum. 

Using conservation of energy \cite{bhagat2018origin} showed that the radial velocity gradient is also singular, in this case $\frac{dh}{dr}\to \infty$, at a critical Weber number which is numerically slightly different. This radius was identified in experiments as a jump in the flow depth to a thicker and slower flow downstream, and excellent quantitative agreement was found in the predicted and observed values of the jump radius. 

There is a small numerical difference between the two predictions of the jump radius $R$ given by conservation of energy and momentum, respectively, namely

\begin{equation*}
    \frac{R}{R_{\textrm{ST}}} = \left( \frac{1}{f'(0)(2\pi)^3} \frac{C_2}{C_1^3} \right )^{1/4} = 0.2705 \, \, \textrm{and} \, \left( \frac{1}{f'(0)(2\pi)^3} \frac{C_3}{C_1^3} \right )^{1/4} = 0.2481,
\end{equation*}
These numerical values are obtained from Watson's similarity profile which, as acknowledged, is only an approximation to the flow in the film, and the predictions are both smaller than the experimentally measured values of 0.289 $\pm$ 0.015. Since $C_1$ is the area under the curve $f(\eta)$ then this will be smaller for the real profile (figure~\ref{fig:Vel_profile}), leading to a larger prediction. Also, since $0\leq\eta\leq 1$, $C_2 > C_3$, the jump radius estimate from momentum conservation is always smaller than that obtained from energy conservation, suggesting that energy is dissipated in the jump.

\vspace{-12pt}
\section*{{Acknowledgement}}
\setlength{\belowcaptionskip}{-10pt} RKB wishes to thank his PhD supervisor Prof Ian Wilson for his advice and support.

\appendix{}
\section{Scaling relation including gravity and surface tension}
\cite{bhagat2018origin} provided a scaling relationship for the jump radius {$R_{\textrm{ST}}$} from the following three conditions: (1) that the radial flow {velocity $u$ and depth $h$} is balanced by viscous drag, $u/R \sim \nu/{h^2}$, (2) continuity $uRh \sim Q$, and (3) the jump is surface tension dominated, which implies at the jump, {the Weber number }$We = \rho u^2 h/\gamma \approx 1$. {Further, their theoretical analysis including gravity gave the condition of hydraulic jump to be
\begin{eqnarray}\label{eq:jump_cond_g}
\frac{1}{We} + \frac{1}{Fr^2} & = & 1,
\end{eqnarray}
where the Froude number $Fr = u/\sqrt{gh}$}.
Incorporating both gravity and surface tension {through the use of (\ref{eq:jump_cond_g})} modifies the scaling relation (\ref{eq:R_ST}) in the form,     
\begin{eqnarray}
R & \sim & R_{\textrm ST}\left[\sqrt{\left(\frac{Q_C}{Q}\right)^2 + 2\left(\frac{Q_C}{Q}\right)} - \left(\frac{Q_C}{Q}\right)\right]^{1/4}.
\label{eq:SC1a}
\end{eqnarray}
As discussed in \S\ref{sec:DA}, this gravitational correction to the radius $R_{\textrm{ST}}$ determined by surface tension alone is 5\% for $Q = 2Q_C$ and 23\% for $Q=10Q_C$.

\appendix{}
\section*{Appendix B. Self-similar velocity profile}
\label{Appendix-1}
{

\label{self similar}
Following \cite{watson1964radial}'s analysis, the conventional literature on the hydraulic jump (e.g. \cite{ bohr1993shallow,bush2003influence, kasimov2008stationary}) assumes a self-similar velocity profile of the form
 \begin{equation*} \label{eq:u1}
 u(r, z) = u_s(r) f(\eta), \; \; \eta \equiv \frac{z}{h(r)}, \; 0 \leq \eta \leq 1,
 \end{equation*}
 where $u_s$ is the surface velocity and $f(0) = 0, f(1) = 1, f^\prime (1) = 0$.
 which further implies (see\cite{watson1964radial}) that 
 \begin{equation*} \label{eq:u2}
 \ub = u_s(r) f(\eta)\hat{r} + u_s(r)h^\prime(r)\eta f(\eta)\hat{z}
 \end{equation*}
 Here we will show that the conventional self-similar velocity profiles do not satisfy the zero interfacial shear-stress condition or $(\nb.\nabla)\ub \neq 0$. We write
 \begin{equation*}
   (\nb.\nabla)\ub = (-\sin\alpha \frac{\partial}{\partial r} + \cos\alpha \frac{\partial}{\partial z})(u_s(r) f(\eta)\hat{r} + u_s(r)h^\prime(r)\eta f(\eta)\hat{z})  
 \end{equation*}
 or,
 \begin{equation*}
     \begin{split}
      [-\sin\alpha \frac{du_s}{dr} f(\eta) -\sin\alpha u_s \frac{h^\prime}{h}\eta f^\prime(\eta) + \frac{1}{h} \cos\alpha u_s f^\prime(\eta)]\hat{r} +   
      [-\sin\alpha h^\prime \frac{du_s}{dr} \eta f(\eta)  -\sin\alpha h^{\prime \prime} {u_s} \eta f(\eta) + \\ \sin\alpha h^{\prime} {u_s} \eta f(\eta) \frac{h^\prime}{h} + \sin\alpha h^{\prime} {u_s}\frac{h^\prime}{h} \eta^2 f^\prime(\eta)  + \cos\alpha{u_s}\frac{h^\prime}{h}f(\eta) + \cos\alpha{u_s}\frac{h^\prime}{h}\eta f^\prime(\eta) ]\hat{z}
     \end{split}
     \label{0stress2}
 \end{equation*}
 Evaluating (\ref{0stress2}) at the free surface with boundary conditions, $f(1) = 1$ and $f^\prime(1) = 0$ yields, 
 \begin{equation*}
     \begin{split}
      [-\sin\alpha \frac{du_s}{dr} ]\hat{r} +   
      (-\sin\alpha h^\prime \frac{du_s}{dr} -\sin\alpha h^{\prime \prime} {u_s}  + \sin\alpha h^{\prime} {u_s} \frac{h^\prime}{h}  + \cos\alpha{u_s}\frac{h^\prime}{h} )\hat{z}
     \end{split}
     \label{0stress2}
 \end{equation*} 
 or, 
 \begin{equation*}
     \begin{split}
      - \frac{h^\prime}{(1+ h^{\prime2})^{1/2}} \frac{du_s}{dr} \hat{r} +   
      (- \frac{h^{\prime2}}{(1+ h^{\prime2})^{1/2}}  \frac{du_s}{dr} - \frac{h^\prime h^{\prime \prime} {u_s}}{(1+ h^{\prime2})^{1/2}}   +  \frac{1}{h} \frac{h^{\prime3} {u_s}}{(1+ h^{\prime2})^{1/2}}   \\ +  \frac{1}{h}\frac{h^\prime}{(1+ h^{\prime2})^{1/2}}{u_s} )\hat{z}
     \end{split}
     \label{0stress3}
 \end{equation*} 
 The analysis implies that only for a completely flat film for which $h^\prime = 0$ and $\nb = \hat{z}$, $(\nb.\nabla)\ub = 0$ can be trivially satisfied. In all other cases the tangential stress is non-zero.   
}

\bibliographystyle{jfm}

\bibliography{jfm-instructions}

\end{document}